\documentclass[english,prd,twocolumn,superscriptaddress,nofootinbib,preprintnumbers]{revtex4-1}
\usepackage[latin1]{inputenc}
\usepackage{amssymb}
\usepackage{color}
\usepackage{float}
\usepackage{amsmath}
\usepackage{amsfonts}
\usepackage{dcolumn}
\usepackage{hyperref}

\usepackage{caption}
\usepackage{subcaption}

\usepackage{tabularx}
\usepackage{graphicx}
\makeatother

\usepackage{babel}
\makeatother

\begin{document}

\title{Bounce and Collapse in the Slotheonic Universe }

\author{Debabrata Adak}
\affiliation{Government General Degree College, Singur, Hooghly-712409.}
\author{Amna Ali}
\affiliation{S. N. Bose National Centre for Basic Sciences, Block JD, Sector-III, Salt Lake, Kolkata 700106, India.}

\begin{abstract}
In this paper, we examine the cosmological dynamics of
 a slotheon field in a linear potential. The slotheon correction term $\frac{G^{\mu\nu}}{2M^2}\pi_{;\mu}\pi_{;\nu}$ respects the galileon symmetry in curved space time. 
 We demonstrate the future evolution of universe in this model. We show that in this scenario, the universe 
 ends with the Big Crunch singularity like the standard case. The difference being that the time at which the singularity
 occurs is delayed in the slotheon gravity. The delay crucially depends upon the strength of slotheon correction. 
 We use observational data from Type Ia Supernovae, Baryon Acoustic Oscillations, and H(z) measurements to 
 constrain the parameters of the model for a viable cosmology,
providing the corresponding likelihood contours. 
\end{abstract}


\maketitle
\section{Introduction}
From the time when scientist have proved that the Big-Bang is the most 
plausible theory to describe the beginning of the universe,
the scientific exploration of the ultimate fate of the universe has grabbed 
attention. Within the framework of Einstein's theory of
general relativity it has been shown that, the fate of the universe consisting 
of pressure less dust depends on its spatial geometry.
A matter dominated universe will expand forever if its spatial geometry is 
hyperbolic or will eventually re-collapse if the spatial geometry is that 
of a three sphere. From the cosmological observations of Supernovae Ia 
(SnIa)\cite{perl} and Cosmic Microwave Background (CMB)\cite{sperg}, 
it is evident
that this simple picture is not true as the universe consists of an additional 
mysterious component, other than radiation and matter. The fate of universe with this additional component is recently investigated 
in literature\cite{ODIN} and has been found that universe generally ends with a collapse. This additional component is in form of an exotic
perfect barotropic fluid with large negative pressure,
dubbed dark energy, which accounts for a repulsive effect
causing acceleration\cite{DE1,DE2,DE3}.  
The recent cosmic acceleration is perhaps the most interesting
phenomenon and nevertheless a very challenging task to the cosmologists
to reveal the reason behind it. Lots of models exist in the literature
which try to describe this phenomenon, however till this date we are unable to 
reach any conclusion.

Initially, right after the discovery of this phenomenon,
the reason behind this cosmic acceleration was thought to be the presence
of cosmological constant in the universe. It is the most simple and 
consistent theory which fits the observations 
very well but is plagued with number of theoretical problems, such as the fine 
tuning and the coincidence problem.
To address these problems, alternative dynamical models of dark energy 
were proposed, one of which are
the scalar field models\cite{scal,scal2,scal3,scal4}. Though scalar fields too are not free from the 
problems associated with the cosmological problems yet some 
 models having
generic features, like the trackers are capable of alleviating
the problems. The major difficulty with the scalar field models are that, a large number of such models are permissible 
by the observational data which makes it difficult to actually 
pin point the actual reason behind the current phenomena of cosmic acceleration.
One must therefore wait for the future observational data which might 
eliminate some of these models and narrow down the
class of permissible scalar field dark energy models.

The other approach to advocate the present cosmic acceleration is the
infra-red modification of the gravity {\it i.e.,} the modification of the 
gravity on the large scales.  
The fact that the quantum mechanical corrections of gravity at the small 
scales are beyond the observational reach at the present day, indicates the 
possibility that the gravity may also suffer modifications at the large scales
where it is not possible to  test it directly. The modified gravity models
have already been proposed on the phenomenological grounds \cite{mod-grav}. 
Moreover
these modification can also arise as the effects of the existence of the 
higher dimensions in the universe \cite{hd}. Building an alternate theory of gravity is a tough task as the viable theory
should be free from the negative energy instabilities such as ghost or 
tachyon instability. Also the theory should be close to $\Lambda CDM$ 
yet should be distinguishable from it. 

The galileon theories are one of such alternate theories of gravity which  
arises at the decoupling limit of Dvali-Gabadadze-
Porrati (DGP) model \cite{dgp}. The galileon theories are a subclass of scalar 
tensor theories which involve only up to second order derivatives as a result
the ghosts do not appear in these theories. These features were originally
found in the Horndeski theory \cite{horn}. The Lagrangian of the galileon 
field $\pi$ respects the shift symmetry in the flat spacetime given by,
\begin{eqnarray}
\pi\rightarrow \pi+ a+ b_\mu x^\mu
\end{eqnarray}
where $a$ is a constant and $b_\mu$ is a constant vector. The present and the future cosmological implication of galileon action has been 
extensively studied in the literature\cite{gali1,gali2,gali3}.
Recently the galileon theory has been generalised to the curved spacetime
\cite{germani1}, such that the modified shift symmetry is given by,
\begin{eqnarray}
\pi(x) \rightarrow \pi(x)+ c + c_\mu \int_{{\cal C},x_1}^{x_2}  \xi^\mu\,\,,
\label{ss2}
\end{eqnarray}
where $\xi^\mu$ is a given set of killing vector and $x_1$ and $x_2$ are two
reference points connected by curve ${\cal C}$. $c$ is a constant and $c_\mu$
is a constant vector.
It is shown  that the Lagrangian $L=-\frac{1}{2}g^{\mu\nu}\partial_\mu \pi
\partial_\nu \pi +\frac{G^{\mu\nu}}{2M^2}\partial_\mu \pi \partial_\nu \pi$
respects this shift symmetry\cite{germani1} and in the corresponding  scalar
field in the flat space time limit  moves slower than that in the  canonical theory. This is solely due to the extra 
gravitational interaction present in the theory. For this reason the
scalar field $\pi$ is called the ``Slotheon''. The authors of 
\cite{aa} have demonstrated the cosmological dynamics
of slotheon field in a potential and have shown that the slotheon term  gives rise to a viable ghost-free late-time
acceleration of the universe.

In the work \cite{da}
authors have shown that the ``high $f$'' issue of
the pNGB quintessence can actually be resolved if terms like
$\frac{G^{\mu\nu}}{2M^2}\partial_\mu \pi \partial_\nu \pi$ are present in
the theory. The shift symmetry of the pNGB field can be broken by the 
presence of a 5-branes placed in highly warped throats \cite{sp}. 
As a result, the effective potential for the axions are slowly varying 
and can be approximated to a linear potential for the axions.

Motivated by this, here we will explore  the cosmological dynamics of the 
slotheonic 
 scalar field $\pi$ in a linear potential. Linear potential has been used to
explain the late time cosmic acceleration in various literatures\cite{lp}. 
The dynamics of linear potential is such that it is quite
insensitive to the initial conditions and it ends with a collapse of 
universe\cite{collapse}.

The paper is organised as follows. In the next section we describe
the  dynamics of slotheon field for a linear potential in the expanding universe.
The equations are solved numerically starting from the matter dominated era
to the accelerated phase. The future evolution of the scalar fields are
also found to check the bounce and collapse in future.

\section{Scalar field dynamics}

We consider a slotheon field $\pi$ with the action:
\begin{align}
S=\int d^4x \sqrt{-g}\Bigl[\frac{1}{2}\Bigl(M_{\rm {pl}}^2 R &-\Bigl(g^{\mu\nu} -\frac{G^{\mu\nu}}{M^2}\Bigr)\pi_{;\mu}\pi_{;\nu}\Bigr)-V(\pi)\Bigr]\nonumber\\
&+ \mathcal{S}_m\Bigl[\psi_m;e^{2 \beta \pi/M_{\rm pl}} g_{\mu\nu}\Bigr]\,,
\label{1.1}
\end{align}
where $M_{\rm Pl}$ is the Planck mass given by $M_{\rm Pl}=1/{8\pi G}$
and $M$ is a mass scale associated with the slotheon field $\pi$. $R$ is the Ricci
scalar, $\psi_m$ is the matter field which couples to $\pi$ with coupling 
constant $\beta$. Here we consider a linear potential as:
\begin{equation}
 V(\pi)= V_0 \pi
\,\,,\\
\label{pot}
\end{equation}
where $V_0$ is a constant, Varying the above action, we obtain the equation of motions as,
\begin{equation}
 M_{\rm Pl}^2 G^{\mu\nu}= T^{\mu\nu}_{(m)}+T^{\mu\nu}_{(r)}+T^{\mu\nu}_{(\pi)}
\,\,,\\
\end{equation}

\begin{equation}
\box\pi\frac{1}{M^2}\left[\frac{R}{2}\box \pi
- R^{\mu\nu} \pi_{;\mu\nu}\right]-V'(\pi) =-\frac{\beta}{M_{\rm Pl}}T_{(m)}
\,\,,
\end{equation}
where $T^{\mu\nu}_{(m)},\,T^{\mu\nu}_{(r)},\,T^{\mu\nu}_{(\pi)}$ are the
energy momentum tensors for the matter, radiation and the scalar field $\pi$
respectively. The symbol ``;'' denotes the covariant derivative and 
``$\prime$'' 
denotes the derivative with respect to the scalar field $\pi$.

\begin{align}
T_{\mu\nu}^{(\pi)}&=
\pi_{;\mu}\pi_{;\nu}-\frac{1}{2}g_{\mu\nu}(\nabla\pi)^2
-g_{\mu\nu}V(\pi)\nonumber\\
&+\frac{1}{M^2} \Bigl[\frac{1}{2}\pi_{;\mu}\pi_{;\nu}R-2\pi_{;\alpha}\pi_{(;\mu}R^{\alpha}_{\nu)}+\frac{1}{2}\pi_{;\alpha}\pi^{;\alpha}G_{\mu\nu}\nonumber\\
&-\pi^{;\alpha}\pi^{;\beta}R_{\mu\alpha\nu\beta}-\pi_{;\alpha\mu}\pi^{\alpha}_{;\nu}+\pi_{;\mu\nu}\pi_{;\alpha}^{~\alpha}\nonumber\\
&+\frac{1}{2}g_{\mu\nu}[\pi_{;\alpha\beta}\pi^{;\alpha\beta}-(\pi_{;\alpha}^{~\alpha})^2+2\pi_{;\alpha}\pi_{;\beta} R^{\alpha\beta}]\Bigr]\,.
\end{align}
Due to the gravitational interaction of $\pi$ with the space-time curvature, there arises a friction as a result
of which the velocity of the field $\pi$ is less than corresponding velocity of the canonical scalar field with same energy\cite{germani1,germani2}..
This holds true even if we add a potential $V(\pi)>0$. 
Though due to the presence of 
potential the action is not $\pi$-parity invariant, yet it is free from 
Ostrogradsky ghost problem. In a spatially flat  FLRW background, the equations of motion take the form

\begin{align}
3M_{\rm{Pl}}^2H^2 &=\rho_m+\rho_r+\frac{\dot{\pi}^2}{2}+\frac{9H^2\dot{\pi}^2}{2M^2}+V{(\pi)}\,,\label{hub}\\
M_{\rm{Pl}}^2(2\dot H + 3H^2)&=-\frac{\rho_r}{3}-\frac{\dot{\pi}^2}{2}+V(\pi)+\frac{\dot{\pi}^2}{2M^2}\Bigl(2\dot H + 3H^2\Bigr)\nonumber\\
&+\frac{2H\dot{\pi}\ddot{\pi}}{M^2}\,,\label{hub2}\\
-\frac{\beta}{M_{\rm{Pl}}} \rho_m &=\ddot{\pi}+3H\dot{\pi}+\frac{3H^2}{M^2}\Bigl(\ddot{\pi}+3H\dot{\pi}+\frac{2\dot{H}\dot{\pi}}{H}\Bigr)\nonumber\\
&+V'(\pi).\label{eompi}
\end{align}
The equations for the conservation of energy follow from the
$\nabla_\mu T^\mu_\nu(\phi)=\frac{\beta}{M_{\rm Pl}}T_m\nabla_\nu\phi$ and
$\nabla_\mu T^\mu_\nu(\phi)=-\frac{\beta}{M_{\rm Pl}}T_m\nabla_\nu\phi$, where
$\nabla_\mu$ represents the covariant derivative and $T_m=-\rho_m$. The 
equations for the conservation of energy are therefore given by,
\begin{align}
\dot\rho_m+3H\rho_m &=\frac{\beta}{M_{\rm{Pl}}}\dot{\pi} \rho_m ,\\
\dot\rho_r+4H\rho_r &=0
\end{align}
\begin{figure*}\centering
\begin{center}
 $\begin{array}{c@{\hspace{.02in}}c}
\multicolumn{1}{l}{\mbox{}}\\ [-0.5cm] &
        \multicolumn{1}{l}{\mbox{}} \\ [-0.5cm]
        \hspace*{-.4in}
        \includegraphics[scale=.8]{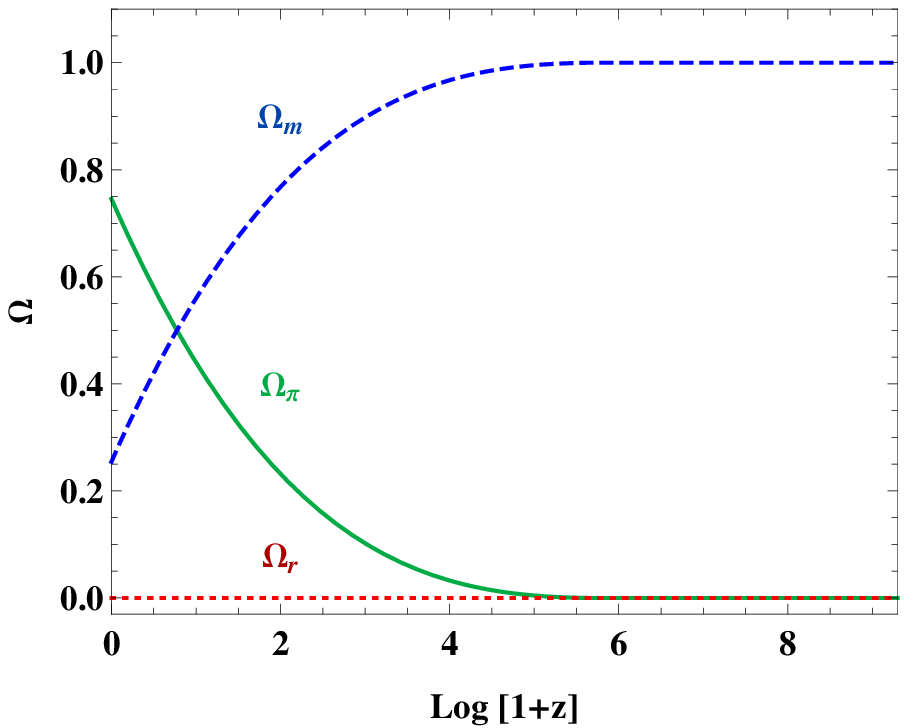} &
                \includegraphics[scale=.82]{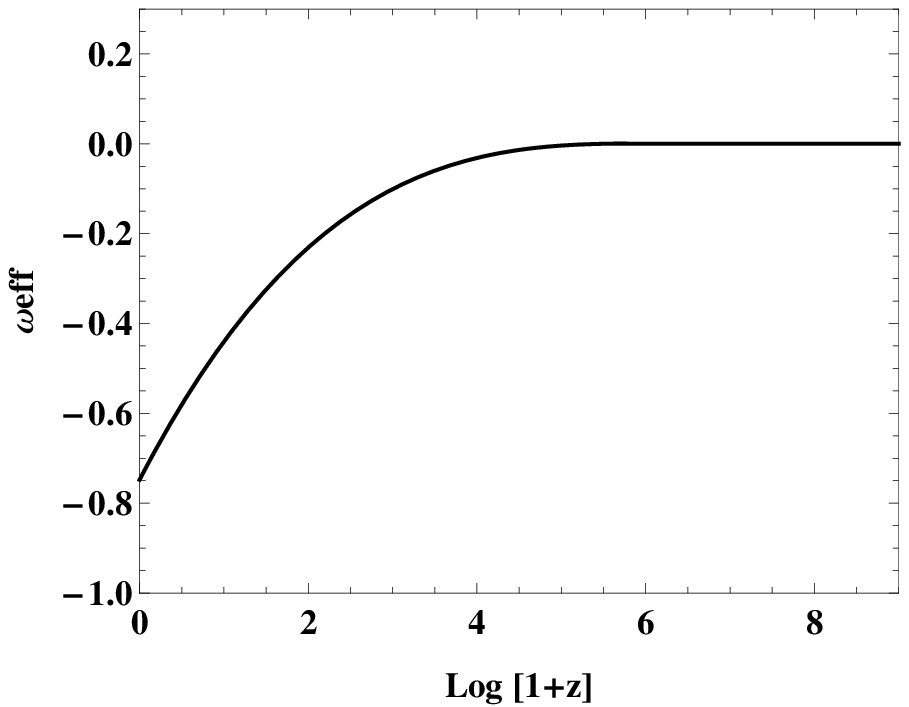} \\ [0.10cm]
\mbox{\bf (a)} & \mbox{\bf (b)}
\end{array}$
\end{center}

\caption{\small The left panel (a):  Density parameters of matter($\Omega_m $), radiation($\Omega_r$) and field($\Omega_{\pi}$) 
for potential (\ref{pot}) are shown here with $V_0=1$ and $\beta=0.01$.
The right panel(b): The cosmic evolution of the  
total effective equation of state $w_\text{eff}$ for the same values of  $V_0$ and $\beta$ is shown. }
\label{omega}
\end{figure*}

$H$ is the Hubble parameter given by
$\dot a/a$ where $a$ is the scale factor of the universe.

The acceleration equation is given by:
\begin{align}
\frac{\ddot a}{a}&=\frac{1}{-2M_{\rm Pl}^2+\frac{\dot\pi^2}{M^2}} \Bigl[M_{\rm Pl}^2 H_0^2\Omega_m^0 a^{-3}+2M_{\rm Pl}^2 H_0^2\Omega_r^0 a^{-4}\nonumber\\
&+\frac{2}{3} \dot\pi+
\frac{\dot\pi^2 H^2}{M^2}-\frac{2}{3}V(\pi)-\frac{2H\dot\pi \ddot\pi}{M^2}\Bigr]\,.
\label{eoma}
\end{align}
$\Omega_m^0$ and $\Omega_r^0$ is the  present density parameter for matter fluid and radiation respectively. Next we define the dimensionless
quantities:
\begin{equation}
 H_0t\rightarrow t_n, \frac{\pi}{M_{\rm Pl}}\rightarrow \pi_n, \frac{V_0}{M_{\rm Pl}H_0^2}\rightarrow V_{0n}, \frac{H_0^2}{M^2}\rightarrow \mu
\end{equation}

In terms of these dimensional quantities the Eqs.\ref{eompi} and \ref{eoma} becomes,
\begin{equation}
 \ddot\pi_n+3H\dot\pi_n+ V_{0n}+3H^2 \mu \Bigl(\ddot\pi_n+3H\dot\pi_n+\frac{2\dot{H}\dot{\pi}}{H}\Bigr)=- 3\beta\Omega_m^0 a^{-3},\, 
 \label{phidot}
\end{equation}

\begin{align}
\frac{\ddot a}{a}&=\frac{1}{(-2+\mu\dot\pi^2)}\Bigl[\Omega_m^0 a^{-3}+2\Omega_r^0 a^{-4}+\frac{2}{3} \dot\pi_n+ H^2\mu\dot\pi_n^2\nonumber\\
&-\frac{2}{3}V_{0n}\pi_n-2H \mu\dot\pi_n \ddot\pi_n\Bigr]
\label{acc}
\end{align}
Here the subscript `n' refers to the new quantities and
the time derivative is taken with respect to $t_n$ . We later drop the subscript `n' for convenience. It is now straightforward to solve these
two equations numerically given the initial conditions.
For this we assume that the universe was matter dominated in the early time. This gives us the following initial conditions:
\begin{eqnarray}
a_{\rm in}&=& \left(\frac{9\Omega_m^0}{4}\right)^{1/3} t_{\rm in}^{2/3}
,\,\nonumber\\
\dot a_{\rm in} &=&  \frac{2}{3}
\left(\frac{9\Omega_m^0}{4}\right)^{1/3} t_{\rm in}^{-1/3},\,\nonumber\\
\pi_{\rm in}&=& \pi_{in},\,\nonumber\\
\dot\pi_{\rm in}&=&0 .\
\label{ini}
\end{eqnarray}
The initial condition for field $\pi_{in}$ is not a free parameter. We  tune it to get the desired present value of matter density$(\Omega_m^0\approx0.3)$ 
and scale factor$(a(t_0)=1)$. With these one can now solve the system numerically.
 In the Fig.\ref{omega}, evolution of the density parameters
of radiation $\Omega_r$, matter $\Omega_m$ and dark energy $\Omega_\pi$ as a function of redshift $z$ are shown. As we have started from matter dominated epoch,
energy density of radiation $\Omega_r$ remains sub-dominant in the entire course of evolution. It is 
quite evident from the figure that the transition from the matter dominated era to the  dark energy 
dominated era takes place recently. Also in this figure we show the evolution of
 the effective equation of state $\omega_{\rm eff}$ $(=-(1+ 2 \dot H/3 H^2))$.  As we start our evolution from matter dominated epoch, the energy 
density of radiation is negligible, therefore $\omega_{\rm eff}=0 $ initially, and  becomes $<-1/3$ when the universe starts
undergoing accelerated expansion.

\section{Future evolution}
We now look for the fate of universe in a slotheonic gravity. From Eqs. \ref{hub} and \ref{hub2}
\begin{figure*}\centering
\begin{center}
 $\begin{array}{c@{\hspace{.02in}}c}
\multicolumn{1}{l}{\mbox{}}\\ [-0.5cm] &
        \multicolumn{1}{l}{\mbox{}} \\ [-0.5cm]
        \hspace*{-.4in}
        \includegraphics[scale=.8]{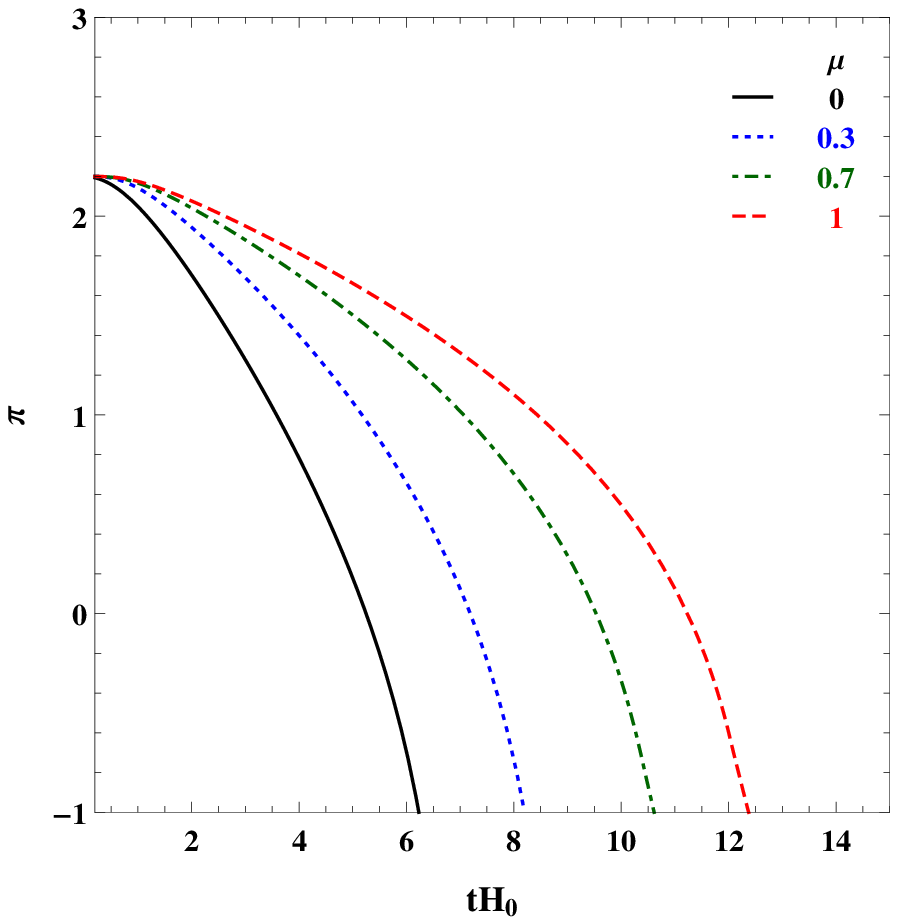} &
                \includegraphics[scale=.82]{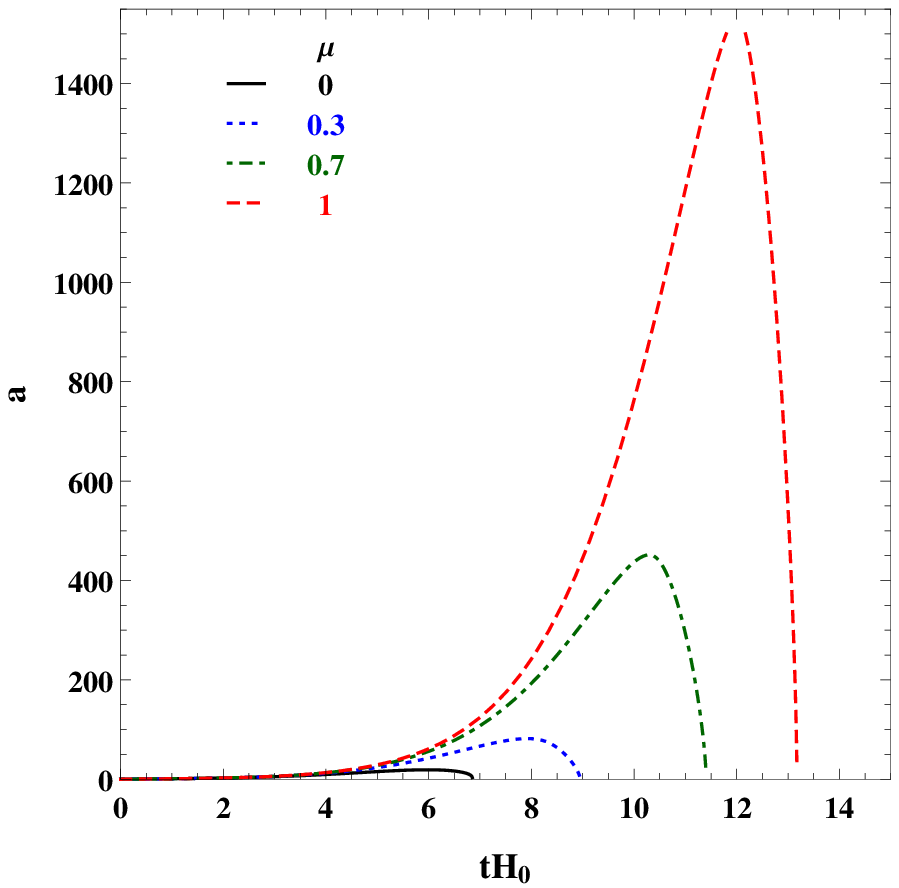} \\ [0.10cm]
\mbox{\bf (a)} & \mbox{\bf (b)}
\end{array}$
\end{center}

\caption{\small The left panel (a) Evolution of the slotheon field $\pi$ for different values of $\mu$ with $V_0=1$ and $\beta=0.01$
The right panel (b)Evolution of scale factor `a' for different values of $\mu$  are shown for the same value of $V_0$ and $\beta$. }
\label{scalefactor}
\end{figure*}
\begin{figure*}\centering
\begin{center}
 $\begin{array}{c@{\hspace{.02in}}c}
\multicolumn{1}{l}{\mbox{}}\\ [-0.5cm] &
        \multicolumn{1}{l}{\mbox{}} \\ [-0.5cm]
        \hspace*{-.4in}
        \includegraphics[scale=.8]{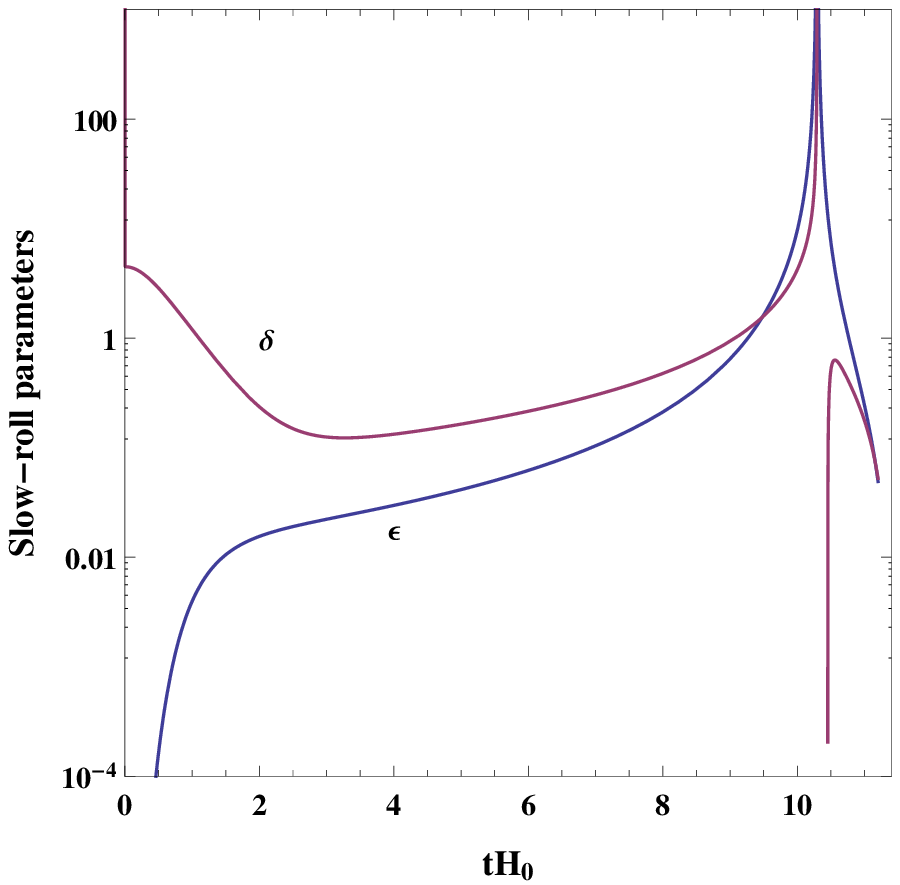} &
                \includegraphics[scale=.82]{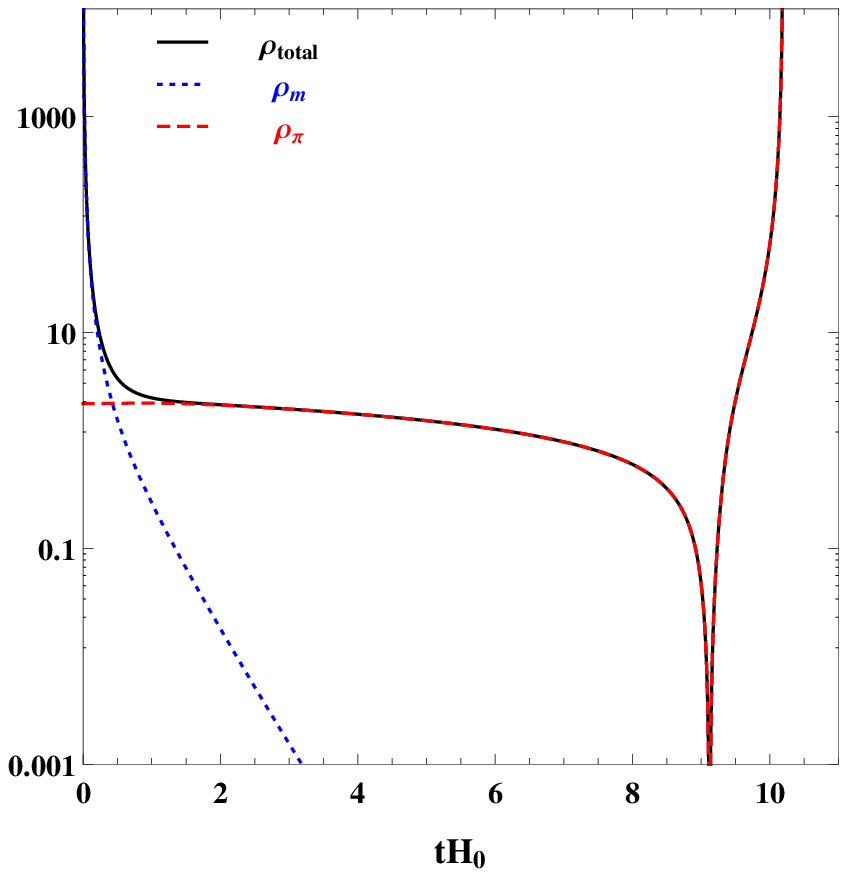} \\ [0.10cm]
\mbox{\bf (a)} & \mbox{\bf (b)}
\end{array}$
\end{center}

\caption{\small The left panel (a) Slow-roll parameters for $V_0=1$, $\mu=0.7$ and $\beta=0.01$.
The right panel (b) Total energy density, matter and field energy density for the same value of $V_0$, $\mu$ and $\beta$. }
\label{slow}
\end{figure*}

From Eqs.\ref{phidot} and \ref{acc}, we notice that when $\mu=0$, the equations
reduces to the standard coupled quintessence field. Therefore the strength of the slotheon gravitation interaction depends on $\mu$.
Extrapolating  Eqs.\ref{phidot} and \ref{acc}, to future such that the present values of density parameters matches the observed values$(\Omega_m^0\approx0.3)$, we get 
the future cosmological dynamics of the slotheon gravity. 
The dynamical evolution of field $\pi$ and the scale factor a for different values of $\mu$ is shown in Fig.\ref{scalefactor}.
Here the present time $t_0$ corresponds to $tH_0=1$. 
We notice that initially,
the positive value 
of field  drives a period of accelerated expansion  but later in future when the field changes the sign, the potential becomes negative, eventually
leading to the collapse of the  scale factor  to a
Big Crunch singularity. We also notice that the change in the sign of field and therefore the Big Crunch singularity depends on the value of $\mu$. For greater $\mu$ the 
collapse of the scale factor is shifted to more distant future. 

As the period of accelerated expansion varies 
with $\mu$, for a given $\mu$ one can determine the length of this period by using the slow-roll parameters $\epsilon\equiv\dot\pi^2/2H^2 M_{\rm Pl}^2$ 
and $\delta\equiv\ddot\pi/H \dot\pi$. For acceleration these two parameters should to be $\ll 1$. Note that this analysis will give us an approximate result
as the slow-roll parameters are only valid in the inflationary paradigm, where the field is the only dominant component.
In the present context, the field begins to evolve in the matter dominated regime, and even at present, the matter content is not negligible.
Though these  traditional slow-roll parameters cannot be
connected to the motion of the field which essentially requires that Hubble
expansion is determined by the field energy density alone, yet it may be helpful to give us a rough idea about the period 
of acceleration. 

In the left panel of fig.\ref{slow} the slow-roll parameters for $\mu=0.7$ is plotted, we notice that both $\epsilon$ and $\delta$
is $\ll 1$ till a period of $\approx9.5 tH_0$ after which it rises steeply signifying the bounce and starting of the collapsing period. It ultimately
collapses at time $\approx 10 tH_0$ which can also be seen from the collapsing of scale factor in fig.\ref{scalefactor}. We also notice that the
second slow-roll parameter $(\delta)$ is not $\ll 1$ around the present time. This is due to the fact that the assumption that they are valid
at present time implies an error in its estimation. Therefore the dynamics is described well at times $t\gg t_0$. The right panel of fig.\ref{slow},
shows the evolution of total energy density $(\rho_{total})$ and energy densities of matter $(\rho_m)$ and field $(\rho_{\pi})$
from early time until the collapse. We see that at early times $\rho_{total}\approx\rho_m$, as matter was dominant but with time as $\rho_m$ decreases, the field dominates, eventually
$\rho_{total}$ becomes equal to $\rho_{\pi}$ around the present epoch. At the time when slow-roll parameters are violated, the field
rolling down the potential reaches a point when $\pi<0$ as a result of which $\rho_{total}$ drops to zero as $H\rightarrow 0$ and a bounce occurs. At this point
of time the other components of universe like matter, radiation or curvature are too less to influence this dynamics.

\begin{figure}
\centerline{\includegraphics[scale=.8]{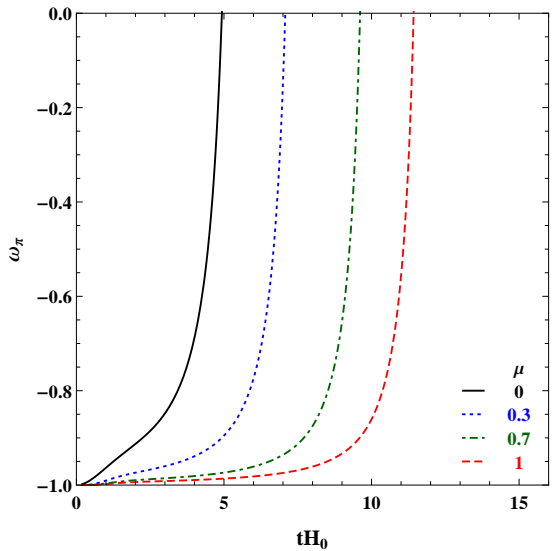} }

\caption{The equation of state of field $(\omega_\pi)$ for different $\mu$ with $V_0=1$ and $\beta=0.01$.}

\label{wpi} 
\end{figure}

The future evolution of equation of state of field $\omega_\pi$ $(=P_\pi/\rho_\pi)$ is shown in fig.\ref{wpi}, we notice that more the strength of the slotheon field, more it 
diverges from the standard coupled quintessence model $(\mu=0)$. 

\section{Observational Constraints on Model Parameters}
In this section, we constrain the parameters of the model with the assumption of a flat Universe by using the latest observational data. 

\begin{figure}
\centerline{\includegraphics[scale=.8]{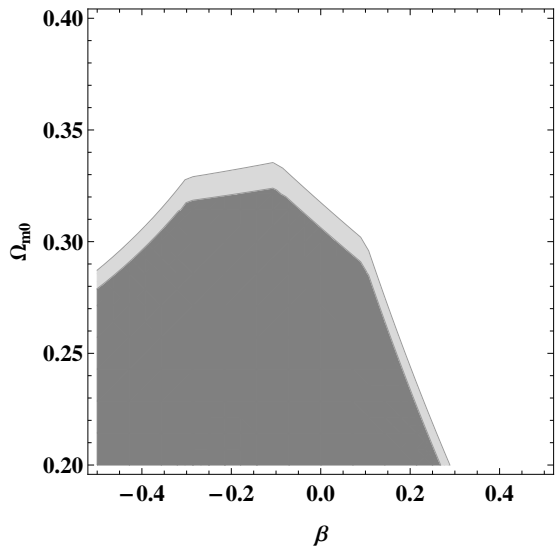} }

\caption{The 1$\sigma$(dark) and 2$\sigma$(light) likelihood
contours in the $(\Omega_{m0} , \beta)$ phase plane for $V_0=1$ and $\mu=0.7$.}

\label{contb} 
\end{figure}

We consider the Supernovae Type Ia observation which
is one of the direct probes for late time acceleration. We
have utilized the Union2.1 compilation of the dataset which
comprises of 580 datapoints~\cite{SnIa}. It measures the apparent brightness of the Supernovae as observed by us which
is related to the luminosity distance
$D_L$  is the luminosity distance defined as
\begin{align}
D_L(z)=(1+z) \int_0^z\frac{H_0dz'}{H(z')}\,,
\end{align}
With this we construct the distance modulus
$\mu_{SN}$,which is experimentally measured:
\begin{align}
 \mu_{SN}= m-M=5 \log D_L+25\,,
\end{align}
 where $m$ and $M$ are the apparent and absolute magnitudes of the Supernovae respectively which are logarithmic measure of flux
and luminosity respectively.
Other observational probe that has been widely used
in recent times to constrain dark energy models is related
to the data from the Baryon Acoustic Oscillations measurements \cite{bao2} by the large scale galaxy survey. In this
case, one needs to calculate the co-moving angular diameter distance $D_V$ 
 as follows:
 \begin{align}
 D_V=\left[\frac{z_{BAO}}{H(z_{BAO})}\left(\int_0^{z_{BAO}} \frac{dz}{H(z)}\right)^2\right]^{\frac{1}{3}}
 \end{align}
 or BAO measurements we calculate the ratio
$\frac{D_V(z=.35)}{D_V(z=.20)}$. This ratio is a relatively model independent quantity and has a measured value  $1.736\pm0.065$.
\begin{figure}
\centerline{\includegraphics[scale=.8]{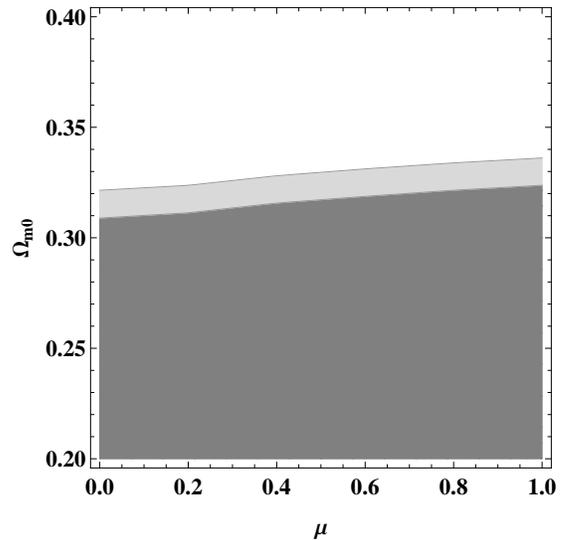} }

\caption{The 1$\sigma$(dark) and 2$\sigma$(light) likelihood
contours in the $(\Omega_{m0} , \mu)$ phase plane for $V_0=1$ and $\beta=0.01$.}

\label{contmu} 
\end{figure}
Next we use Hubble data from red-envelope galaxies. 12 measurements of
the Hubble parameter H(z) at redshifts $.2<z<1$ are obtained from a high-quality spectra with the Keck-LRIS
spectrograph of red-envelope galaxies in 24 galaxy clusters\cite{stern}.  The measurement at
$z= 0$ was from HST Key project \cite{hst}.At
this point we define the normalised hubble parameter as
$h(z) =\frac{H(z)}{H_0}$
and utilise it to derive the value of new $h(z)$. Using all these observational data, we constrain the model parameters $\mu$ and $\beta$ to see 
what is allowed by the observational data. In fig.\ref{contmu} we show the confidence contours in the
$(\Omega_{m0} , \mu)$
parameter space. We notice that  $\mu$ is unconstrained by the data. The confidence contours in parameter space $(\Omega_{m0} , \beta)$
is shown in fig.\ref{contb} for $\mu=0.7$. We notice that $\beta$ is  constrained by the data to small
values $\beta < 0.3$.
\section{Conclusion}

In this work we investigated the slotheon gravity in a linear potential. We have shown that cosmological
dynamics of this model is similar to the coupled quintessence model at late times, thereby giving 
an accelerated expansion at recent time. The dynamics of  linear potential is such that triggers a collapse of universe\cite{collapse}. 
The collapse occur when the field moves down 
slowly encounters a negative potential energy. The energy density of 
the universe eventually becomes zero due to which the universe bounces 
and a collapsing
period starts dominating by the kinetic energy of the field. 
Here we  have extended this formalism to the slotheon gravity to study how it is different from the standard quintessence case.  When the slotheon 
gravity strength $\mu=0$, the slotheon field reduces to the standard coupled quintessence. Generally in this case, in an expanding universe,
a scalar field will dominate the energy density around the present epoch and drive a period of cosmic acceleration, followed by a period
of bounce and collapse. The nature of the collapse is that of the Big Crunch singularity.

We have shown that when the slotheon gravity comes into play $(\mu>0)$, the fate of the universe is similar to that of the standard case. The only
difference is the time at which the Big Crunch singularity occurs. It is shown that the collapse is shifted to a distant future in the slotheon
gravity and can be made redundant for large value of parameter $\mu$. We have estimated the time of bounce for a particular value of field strength
$(\mu=0.7)$ using the slow-roll parameters. Though it gives an approximate result, yet it gives a rough idea about this period and subsequent events following it.

We have also constrained the model parameters $\mu$ and $\beta$ by using the observational data from Supernovae Type Ia, BAO and H(z) measurements. We see that all values of slotheon gravity
strength $\mu$ is allowed  whereas small values of the coupling constant $\beta$ is preferred by the data.


\begin{thebibliography}{99}
\bibitem{perl}
S. Perlmutter et al., Astrophys. J. {\bf 517}, 565 (1999);
A. G. Riess et al., Astron. J. {\bf 116}, 1009 (1998).

\bibitem{sperg}
 D. N. Spergel et al. (WMAP Collaboration), Astrophys. J.
Suppl. Ser. {\bf 170}, 377 (2007).

\bibitem{ODIN}
S.D. Odintsov, V.K. Oikonomou, Phys. Rev. D {\bf 94} 064022 (2016);
S. Nojiri, S.D. Odintsov, V.K. Oikonomou, Phys. Rev. D {\bf 93} 084050 (2016);
S.D. Odintsov, V.K. Oikonomou, Phys. Rev. D {\bf 92} 024016 (2015);
S.D. Odintsov, V.K. Oikonomou, Phys. Rev. D {\bf 90} 124083 (2014).
 	

\bibitem{DE1}
V. Sahni and A. A. Starobinsky, Int. J. Mod. Phys. D9, {\bf373}
(2000); S. M. Carroll, Living Rev. Relativity {\bf4}, 1 (2001);
T. Padmanabhan, Phys. Rep. {\bf380}, 235 (2003); P. J. E.
Peebles and B. Ratra, Rev. Mod. Phys. {\bf75}, 559 (2003);
E. V. Linder, J. Phys. Conf. Ser. {\bf39}, 56 (2006);  L. Perivolaropoulos, AIP Conf. Proc. {\bf848},
698 (2006).

\bibitem{DE2}
 E. J. Copeland, M. Sami, and S. Tsujikawa, Int. J. Mod.
Phys. D {\bf15}, 1753 (2006).

\bibitem{DE3}
 E. V. Linder, Gen. Relativ. Gravit. {\bf40}, 329 (2008); J.
Frieman, M. Turner, and D. Huterer, Annu. Rev. Astron.
Astrophys. {\bf46}, 385 (2008).

\bibitem{scal}
R. R. Caldwell, R. Dave, and P. J. Steinhardt, Phys. Rev. Lett. {\bf 80},1582 (1998);
T. Chiba, T. Okabe, and M. Yamaguchi, Phys. Rev.
D {\bf 62}, 023511 (2000).

\bibitem{scal2}
I. Zlatev, L. M. Wang, and P. J. Steinhardt, Phys. Rev. Lett.
{\bf82}, 896 (1999); P. J. Steinhardt, L. M. Wang, and I. Zlatev,
Phys. Rev. D {\bf59}, 123504 (1999); L. Amendola, Phys. Rev.
D {\bf62}, 043511 (2000).
\bibitem{scal3}
C. Armendariz-Picon, V. Mukhanov, and P. J. Steinhardt,
Phys. Rev. Lett. {\bf85}, 4438 (2000); Phys. Rev. D {\bf63}, 103510
(2001); T. Chiba, T. Okabe, and M. Yamaguchi, Phys. Rev.
D {\bf62}, 023511 (2000).

\bibitem{scal4}
R. Kallosh, J. Kratochvil, A. Linde, E. Linder, and M.
Shmakova, J. Cosmol. Astropart. Phys. {\bf10}, 015 (2003); Y.
Wang {\it et al},
Cosmol. Astropart. Phys.{\bf12}, 006 (2004); P. P. Avelino, C. J. A. P. Martins, and J. C. R. E. Oliveira, Phys. Rev. D
{\bf70}, 083506 (2004); P. P. Avelino, Phys. Lett. B. {\bf611}, 15
(2005).

\bibitem{mod-grav}
S. Capozziello, Int. J. Mod. Phys. D {\bf 11} 483 (2002);
S. Capozziello, S. Carloni and A. Troisi, Recent Res.
Dev. Astron. Astrophys. {\bf 1} 625 (2003);
Shin'ichi Nojiri, Sergei D. Odintsov, Phys. Rept. {\bf 505} 59-144 (2011).

\bibitem{hd}
A. Nicolis, R. Rattazzi, and E. Trincherini, Phys. Rev. D
{\bf 79}, 064036 (2009); C. Deffayet, G. Esposito-Farese, and
A. Vikman, Phys. Rev. D {\bf 79}, 084003 (2009).

\bibitem{dgp}
G. R. Dvali, G. Gabadadze and M. Porrati, Phys. Lett.
B {\bf 485}, 208 (2000).

\bibitem{horn}
G. W. Hordenski, Int. J. Theor. Phys. {\bf 10}, 363-384 (1974).

\bibitem{gali1}
R. Gannouji and M. Sami, Phys. Rev. D {\bf82}, 024011 (2010);
 A. Ali, R. Gannouji, and M. Sami, Phys. Rev. D {\bf82},
103015 (2010).


\bibitem{gali2}
A. Ali, R. Gannouji, M. W. Hossain, and M. Sami, Phys.
Lett. B {\bf718}, 5 (2012).

\bibitem{gali3}
M. Shahalam, S. K. J. Pacif, R. Myrzakulov,
 Eur. Phys. J. C {\bf76}, 410 (2016). 


\bibitem{germani1}
C. Germani, L. Martucci, and P. Moyassari, Phys. Rev. D
{\bf 85}, 103501 (2012).

\bibitem{germani2}
C. Germani, Phys. Rev. D {\bf 86}, 104032 (2012).




\bibitem{da}
D. Adak and K. Dutta, Phys. Rev. D{\bf 90}, 043502 (2014). 

\bibitem{aa}
D. Adak, A. Ali and D. Majumdar, 
Phys. Rev. D{\bf 88}, 024007 (2013).

\bibitem{sp}
Sudhakar Panda, Yoske Sumitomo, Sandip P. Trivedi,
Phys. Rev. D{\bf 83}, 083506 (2011).

\bibitem{lp}
Robert J. Scherrer and A. A. Sen,
Phys. Rev. D {\bf 77}, 083515 (2008);
L. Perivolaropoulos
Phys. Rev. D {\bf 71}, 063503 (2005).


\bibitem{collapse}
Ricardo Z. Ferreira and Pedro P. Avelino,
arXiv:1508.00631 [astro-ph.CO].

\bibitem{SnIa}
N. Suzuki {\it et al},
Astrophys. J. \textbf{746} (2012) 85.

 
\bibitem{bao2} 
 W. J. Percival {\it et al},
 Mon. Not. Roy. Astron. Soc. \textbf{401}, 2148 (2010).
 
 \bibitem{stern}
D. Stern, R. Jimenez, L. Verde, M. Kamionkowski and
S. Adam, 
JCAP \textbf{1002},(2010) 008. 

\bibitem{hst}
W. L. Freedman et al. [HST Collaboration],  Astrophys. J. \textbf{553} (2001)
47.
  	


\end{thebibliography}
\end{document}